# Modeling Uncertainty and Evolving Self-Adaptive Software: A Fuzzy Theory Based Requirements Engineering Approach

Zhuoqun Yang[1], Zhi Jin[2], Zhi Li[3]

[1]Institute of Mathematics, Academy of Mathematics and Systems Science, Chinese Academy of Sciences,

Haidian Dstr., Beijing 100190, P. R. China

[2]Key Laboratory of High Confidence Software Technologies (MoE), Peking University,

Haidian Dstr., Beijing 100871, P. R. China

[3]Software Engineering Dept., College of Computer Science and Information Technology, Guangxi Normal University,

Guilin, Guangxi 541004, P. R. China

**Abstract**
Self-adaptive software (SAS) is capable of adjusting its behavior in response to meaningful changes in the operational context and itself. Due to the inherent volatility of the open and changeable environment in which SAS is embedded, the ability of adaptation is highly demanded by many software-intensive systems. Two concerns, i.e., the requirements uncertainty and the context uncertainty are most important among others at Requirements Engineering (RE) stage. However, requirements analyzers can hardly figure out the mathematical relation between requirements, system behavior and context, especially for complex and nonlinear systems, due to the existence of above uncertainties, misunderstanding and ambiguity of prior knowledge. An essential issue to be addressed is how to model and specify these uncertainties at RE stage and how to utilize the prior knowledge to achieve adaptation.

In this paper, we propose a fuzzy-based approach to modeling uncertainty and achieving evolution. The approach introduces specifications to describe fuzziness. Based on the specifications, we derive a series of reasoning rules as knowledge base for achieving adaptation and evolution. These two targets are implemented through four reasoning schemas from a control theory perspective. Specifically, forward reasoning schema is used for direct adaptation; backward reasoning schema is used for optimal adaptation. Parameter-identified schema implements learning evolution by considering SAS as the gray-box system, while system-identified reasoning schema implements learning evolution by considering SAS as the gray-box system. The former two schemas function as the control group, while the latter two are designed as the experimental groups to illustrate the learning ability. Our approach is implemented under three types of context: derivable, quasi-noisy and noisy context, through the demonstration of a mobile computing application.

By comparing the accuracy performance and time performance of these schemas, we show the availability in modeling and reasoning over uncertainties. It also shows the inconsistencies between users' preferences and prior knowledge at RE stage, which can be diminished through learning with our approach. Besides, our approach can be used as a solution to reasoning over continuously variables.

**Keywords**    self-adaptive software, requirements engineering, uncertainty, evolution, fuzzy theory

## 1. Introduction

The self-adaptive software (SAS) system is a novel computing paradigm in which the software is capable of adjusting its behavior in response to meaningful changes in the environment and itself [1]. The ability of adaptation is characterized by self-*properties, including self-healing, self-configuration, self-optimizing and self-protecting [2]. Innovative technologies and methodologies inspired by these characteristics have already created avenues for many promising applications, such as mobile computing, ambient intelligence, ubiquitous computing, etc.

Software-intensive systems are systems in which software interacts with other software, systems, devices, sensors and with people intensively. Such an operational environment may be inherently changeable, which makes self-adaptiveness become an essential feature. Context can be defined as the reification of the environment [3] that is whatever provides as a surrounding of a system at a time. It provides a manageable and manipulable description of the environment. Context is essential for the deployment of self-adaptive software. As the environment is changeable, the context is unstable and ever changing and the system is desired to perform different behaviors according to different contexts. Therefore, engineers need to build effective adaptation mechanisms to deal with contextual changes.

Requirements Engineering (RE) for self-adaptive systems primarily aims to identify adaptive requirements, specify adaptation logic and build adaptation mechanisms [4]. Conducting context analysis at the requirements phase will be worthwhile at the design and development phases, because contexts may influence the decisions about what to build and how to build them. However, some kinds of uncertainty may occur in both context and requirements [5]. First, it is often infeasible to precisely detect, measure and describe all the contextual changes. This imprecision about how contextual changes at runtime can be viewed as context uncertainty [6]. Second, the extent to which the non-functional requirements (NFRs) are satisfied, and the task configurations, w.r.t. functional requirements (FRs), with which the system operates in changing contexts, are also uncertain. These kinds of uncertainties are known as requirements uncertainties [7]. By saying requirements uncertainties, it means we do not know how NFRs changes, but we can map contextual changes to changes of NFRs. Once contexts are fixed, at a certain time point, NFRs are fixed according to user preferences. Similarly, task configurations can also be determined after contexts are monitored. Dealing with these two kinds of uncertainties becomes a challenge for the research community of RE for SAS.

Many research works in the literature have shown remarkable progress in providing solutions to mitigating these uncertainties. **For dealing with requirements uncertainty**, a research agenda is provided in [8]. The author argues requirements for self-adaptive systems should be viewed as runtime entities that can be reasoned over in order to understand the extent to which they are being satisfied and to support adaptation decisions that can take advantage of the systems' self-adaptive machinery. More recently, related works are fully synthesized and summarized in a roadmap paper [6]. Other works from the viewpoint of RE focus on modeling and specifying aspects. FLAGS [9] is proposed for mitigating the requirements uncertainty by extending the goal model with adaptive and fuzzy goals. It provides a general overview of how these goals can be handled as runtime abstractions and specifies the extent to which tasks are operated, when requirements can be satisfied. However, they do not figure out how to tune tasks in order to satisfy requirements. RELAX [10] is a formal requirements specification language, which is defined in terms of temporal fuzzy logic. It is introduced to relax the objective of SAS. Different from FLAGS, RELAX not only describes relaxed requirements, but also captures the environment. Through this approach shows lots of advantages in identifying uncertainty with fuzzy logic and establish the boundaries of adaptive behavior, it is limited to reasoning with the specification and computing the adaptation decision. For coping with uncertainties of system configurations, some works proposed approaches of adaptation at the design stage. FUSION [11] uses online learning to mitigate the uncertainty associated with changes in context and reconfigure features from feature pool to unanticipated changes. The method utilizes linear equations to simulate the system for reducing time complexity of decision-making. However, for most of software systems, building the analytic formulae of context and system is a tough nut and highly depends on analyzers' domain knowledge. POISED [12] improves the quality attributes of a software system through reconfiguration of components to achieve a global optimal configuration for the software system. The adaptation of this approach is also based on existing finite configuration options. However, at the requirements phase, we can hardly capture the precise and appropriate configuration candidates. **For describing contexts**, Ali et al. [13, 14] conduct remarkable works by proposing a goal-oriented RE modeling and qualitative reasoning framework for systems operating in varying contexts. It introduces contextual goal models to relate goals and contexts and reasoning techniques to derive requirements reflecting the context and users priorities at runtime. However, their approach is not appropriate for specifying and reasoning with the continually changed contexts, because most of time, the context in which SAS is deployed has a continuous value, e.g. the temperature of a room and the intensity of a lamp. This kind of variability raises the difficulties of modeling and reasoning with contexts.

Except for the above research gaps, two difficulties should be addressed. First, NFRs may evolve according to the contextual changes and the evolution may modify the criteria on which the trade-off of adaptation decision is based. The mappings from contexts to NFRs and from system tasks to NFRs depend on stakeholders' preferences [15], while the mapping from context to system tasks is based on domain knowledge. We cannot confirm the consistency between user preference and domain knowledge during trading off NFRs when domain knowledge is invalid for reuse. This situation demands the ability of learning knowledge and evolution at runtime. Second, due to the informal nature of RE activities, e.g., the inherent fuzziness and vagueness of human perception, understanding and communication of their desire in relation to the non-formal real world, we cannot precisely define the analytic formulae between changing contexts and the system requirements [16]. This situation demands the ability of computing with uncertainties.

All these research challenges urgently demand innovations of the existing approaches and techniques at RE phase for modeling uncertainty, reasoning with uncertainty, achieving adaptation and evolution. In this paper, uncertainties are described as fuzzy attributes of modeling elements. Based on the specification of fuzziness, we derive a series of reasoning rules as the knowledge base for achieving adaptation and evolution. Though reasoning with fuzziness attributes, SAS can achieve the optimal configuration (adaptation) and continually learn knowledge at runtime (evolution) according to stakeholders' preferences and system boundaries. To this end, we design four reasoning schemas from the perspective of control theory. Specifically, forward reasoning schema is used for achieving direct adaptation, which illustrates the inadequacy of prior knowledge at runtime; backward reasoning schema is use for generating the optimal reconfigurations. These two schemas function as the control group. Parameter-identified reasoning schema is implemented by viewing SAS as gray-box system and learns parameters of membership functions according to contextual changes. System-identified reasoning schema is implemented by considering SAS as a black-box system and learns analytic formulae of contexts and configurations. The latter two schemas are designed as the experimental group. For validating our approach, we operate experiments on a mobile computing application.

Our contributions are multifold. First is that we propose a general treatment for modeling SAS, specifying and reasoning over uncertainties, achieving adaptation and evolution from the viewpoint of requirements engineering. Compared with the existing research, our approach is more appropriate for reasoning and decision-making at RE stage when there are only elicited user preferences and prior domain knowledge. Second, we propose how to map the specification of uncertainties to reasoning rules and design flexible reasoning schemas to achieve adaptation and evolution. The reasoning processes of our approach can be applied to continuous variables, which make the adaptation more precise. Third, the adaptation is derived by solving the multi-objective decision-making problem. At each time point, the system configuration is decided according to the trade-off among NFRs. At last, we propose how to deploy a learning mechanism into SAS. With the learning mechanism, SAS can evolve itself with incremental knowledge. Meanwhile, we show that the system gets flexible with the enriched knowledge base.

The paper is structured as follows. Section 2 briefly introduces the basic concepts used in our work. Section 3 provides the motivating example and the overall approach, in order to deliver an easy understanding of this paper. The concepts, models and specifications for describing uncertainties are presented in Section 4. Section 5 provides how to generate reasoning rules based on the specifications. Section 6 elaborates the four types of reasoning schemas and corresponding algorithms. The evaluation of our approach is presented in Section 7. Related works are discussed in Section 8, followed by conclusion and future work in Section 9.

## 2. Background

This section gives a brief introduction to the concepts used in the modeling for self-adaptive software and reasoning with uncertainties. It also describes the relations between these concepts and our work.

### 2.1 Goal Model

Goal model and the goal-oriented analysis are proposed in the RE literature to present the rationale of both humans and systems. A goal model describes the stakeholder's needs and expectations for the target system. KAOS model [17] is one of the most significant goal-oriented requirements model. The core elements of KAOS model include goals, tasks and softgoals. Goals model stakeholders' intentions while the tasks model the functional requirements which can be used to achieve the goals. Goals can be refined through AND/OR-decompositions into sub-goals or can be achieved by sub-tasks. For AND-decomposition, a parent goal will be satisfied when all its sub-elements are achieved, while for OR-decomposition, a parent goal can be satisfied by achieving at least one of its sub-elements. OR-decompositions incorporate and provide sets of alternatives which can be chosen flexibly to meet goals. Softgoals model the NFRs, which have no clear-cut criteria for their satisfaction and can be used to evaluate different choices of alternative tasks. Tasks can contribute to softgoals through the Help/Hurt contribution relation. In our work, we adopt the modeling elements of KAOS model to constructing requirements model. For modeling relations between uncertain entities, we extend the initial Help/Hurt relations with new semantics.

### 2.2 Fuzzy Reasoning Machine

Fuzzy control [18] is a practical alternative for achieving high-performance control on nonlinear time-variant system since it provides a convenient method for constructing nonlinear controllers using heuristic rules. Heuristic rules may come from domain knowledge. Engineers incorporate these rules into a fuzzy controller that emulates the decision-making process of the human. A fuzzy controller has four principal components: (1) *Rule base* holds the knowledge in the form of a set of control rules, of how best to control the system. (2) *Fuzzification* process modifies the crisp input with membership functions and the output is known as membership degree, so that they can be interpreted and compared according to the rules in the rule base. (3) *Reasoning machine* evaluates which control rules are relevant to the current input and then decides what the membership degree of output should be. (4) *Defuzzification* process converts the conclusions in the form of membership degree into the crisp output. A set of membership functions is responsible for all the transforming processes. In our work, we apply fuzzy reasoning mechanisms for each reasoning process. Fuzzy reasoning mechanism ensures reasoning over uncertainties and continuous variables.

### 2.3 Learning Algorithm

Genetic Algorithm (GA) [19] is a kind of heuristic algorithm for implementing optimization problems. The variables are mapped to chromosome of populations. After the selection, crossover, mutation and recombination of chromosome, we can derive the optimal value. Meanwhile, the corresponding values of independent variables are the optimal solution to the problem. Neural Networks (NNs) [20] is an information processing paradigm that is inspired by the way biological nervous systems, such as the brain, process information. The commonest type of NNs consists of three layers: input layer, hidden layer and output layer. It can solve the optimization problem by tuning the weight between these layers. We apply GA into the optimization process of BR schema and the parameter-identification process of PR schema. NNs are used for implementing the system-identification in SR schema.

## 3. Approach Overview

In this section, we first present the motivating example. Based on the example, we detail the problems of uncertainty need to be solved. Then, we provide the overall process of our approach to deliver an easy understanding.

### 3.1 Motivating Example

To illustrate the proposed approach, we consider the push notification technology in the mobile computing domain. Typically, push notifications is a technique used by apps to alert smartphone owners on content updates, messages, and other events that users may want to be aware of. This technique has been successfully developed as APIs on iOS systems, such as Prowl (http://www.prowlapp.com) and Pushover (https://pushover.net/). These applications focus on receiving the needed infor-mation timely, no matter where the user is. However, for the location-related scenarios, the application's performance tightly related to the user's location, e.g. learning application on smartphone [21].

The mobile business application is the software deployed on smartphones to support pushing business information to the nearest customers in a smart business district. The objective of pushing business information is to notify users of surrounding information and events, such as the goods on sale, goods' description, comments of customers, etc. Figure 1 presents a requirements model built with KAOS model. To achieve the top goal (g0), the application needs to locate users (g1) and set the receiving configuration (g2). In this paper, we consider four tasks for achieving these two goals. To satisfy goal g1, either task t1 or t2 can be performed. We use them for illustrating structural adaptation [22]. To achieve goal g2, both task t3 and t4 should be completed. We use them for illustrating parametric adaptation. Except for the functional concerns, the application has three NFRs: high time efficiency (sg1) referring to the requirement for a shorter response time and update time interval; high energy efficiency (sg2) referring to the requirement for a higher battery of mobile phone; high information efficiency (sg3) referring to the requirement for better-timed and larger-sized notifications. The related contexts in which the application operates are presents with the contextual model [13] in Figure 2, including bandwidth rate, network delay, dump energy and available memory. The former two belong to external contexts, while the latter two belong to internal contexts.

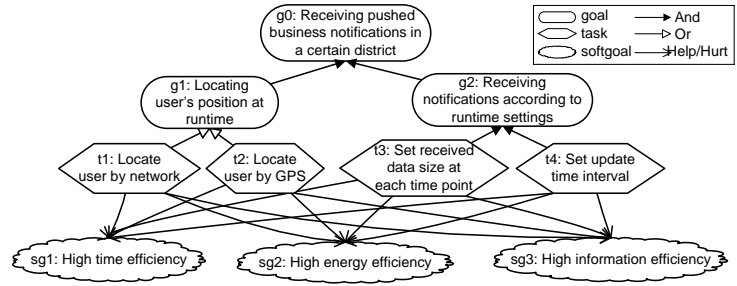

**Figure 1 KAOS model of mobile business application.**

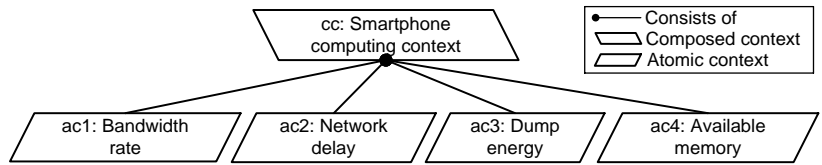

**Figure 2 Context model of mobile business application**

### 3.2 Problem statement

The contexts where mobile business application is deployed are ever changing. Especially, for contexts like bandwidth rate, the changing may be more rapid. Thus, the performance of mobile business application highly depends on contexts. This dependency is implied in two aspects: the satisfaction degrees of NFRs depend on context and task configuration about FRs depends on context.

The former dependency can be elicited according to users' preferences and we refer the satisfaction degrees of NFRs defined by this process to the desired satisfaction degrees. For example, when dump energy of a smartphone is too low (ac3), the user demands the configuration for saving energy and no longer cares how much information he can get. Thus, in Figure 1, the desired satisfaction degree of sg2 upgrades, while the degree of sg3 degrades.

The latter dependency can be built according to domain knowledge. For example, when network delay is larger than 10ms (ac2), we choose to apply GPS to locate every 3 seconds (t2). Once this dependency is determined, the system can directly respond to contextual changes.

Except for these two dependencies, the last dependency we concern is implied in the system itself, the satisfaction degrees of NFRs depending on how well the task performs. This dependency can be defined by users and we refer the satisfaction degrees of NFRs determined by this process to actual satisfaction degrees. For example, when the received data size is set to 50MB (t3), the user is happy to get the abundant information (sg3). On the other hand, receiving this size of data may spend too much time, which becomes an obstacle to user experience on sg1.

Due to the uncertainties lie in contexts, NFRs and system configurations, we refer to the above three dependencies as uncertainty dependencies. The problems about uncertainties we discuss and tackle in this paper are caused by these uncertainty dependencies.

**Description problem**

Description problem refers to how to model and specify these uncertainties and uncertain dependencies. As we discussed in Section 1, there are research gaps in settling this problem. We consider use fuzziness to describe uncertainty, because fuzziness captures the ambiguity in changes of SAS and contexts.

**Decision-making problem**

Decision-making problem means how to achieve the optimal system configurations when contexts change. We consider it as the process to achieve adaptation. Due to user preferences change along with contexts, the actual satisfaction degrees of NFRs may deviate from the desired ones. Once deviations emerge, it implies the unavailability of domain knowledge. Thus, decision-making process optimizes system configurations to satisfy NFRs at runtime.

**Learning problem**

Learning problem refers to how to evolve the system through learning incremental knowledge during decision-making. The initial knowledge base is constructed with domain knowledge. When existing knowledge is unavailable in new contexts, new knowledge are learned and added to the knowledge base. The learned knowledge should be used to guide latter self-adaptation. With the help of learned knowledge, the system performs adaptation more flexibly and efficiently. We show this advantage by comparing the performances of adaptation mechanisms with and without learning ability.

### 3.3 Overall Process

Figure 3 presents the general modeling and evolving processes by using our approach. The process includes four steps:

Modeling and specifying uncertainties is the first step for the whole adaptation and evolution processes. In this step, uncertainties are captured and described within uncertain entities and relations. We present how to describe fuzzy attributes with our illustrating example. All the specified fuzzy attributes can be determined by stakeholders and requirements analyzers.

Generating reasoning rules integrates the specifications of fuzziness into IF-THEN clauses, which can be implemented with fuzzy reasoning machine. Reasoning rules include two types: rules for gray-box system and rules for black-box system, which can be applied to different reasoning schemas.

Designing adaptation mechanism is the most significant step. In this step, four reasoning schemas are designed based on control mechanisms. FR is used to illustrate the deviations between user preferences and domain knowledge, while BR is used to derive optimal configurations. PR and SR are implemented with learning ability. They both perform adaptation and evolution. The difference lies in that the evolution of PR is based on parameter identification, while the evolution of SR is based on system identification, which will be detailed in Section 5. After this step, we derive the adaptation mechanisms and algorithms for target optimization problems.

Decision-making and learning knowledge process provides the performance of our approach based on a series of comparable experiments. In this step, we detail the design of experiments for the mobile business application and conduct detailed data analysis. By comparing these results, we present the validity of our approach and illustrate that SAS indeed learns available knowledge during adaptation.

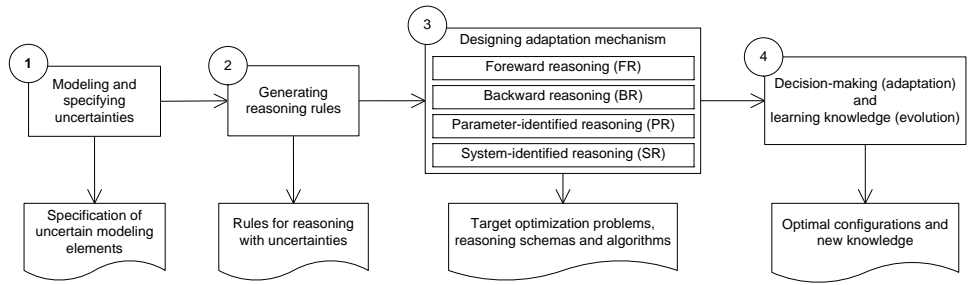

**Figure 3 Modeling and evolving processes for self-adaptive software.**

### 4. Modeling and Specifying

This section first introduces the conceptual model we used in our approach. Then, we provide the definitions and specifications of the modeling entities and relations. Figure 4 presents a generic conceptual model which includes entities and relations considered in modeling and reasoning processes. We adopt the basic modeling elements of KAOS [17] and contextual goal model [13, 14].While, to describe uncertainties and reasoning relations in our approach, we extend the modeling elements and the semantics. These elements are specified in the rest of this section. In this paper, we only focus on the uncertainty attributes of these modeling elements. For the specification of business logic, such as business attributes and adaptation attributes, readers can refer to our previous [23] for more details.

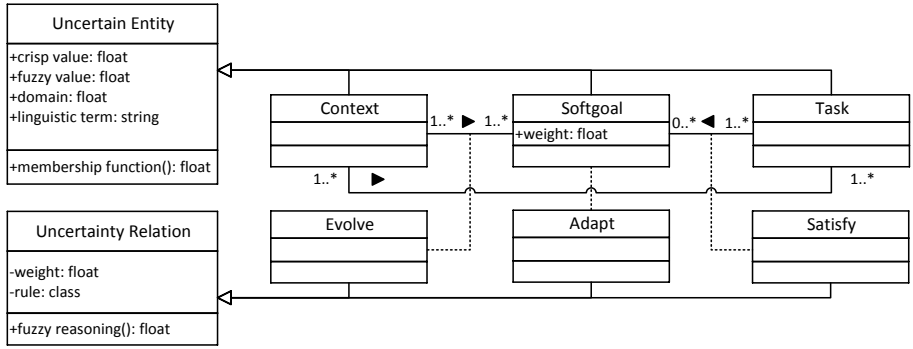

**Figure 4 Concept model for capturing uncertainties.**

## 4.1 Specifying Entity

**Definition 1 (Uncertain Entity)** An atomic uncertain entity is the entity whose value is characterized by linguistic descriptions and variability of the value is independent with other entities of the same type. We specify the entity with $ue := \langle cv, dom, \overrightarrow{lv}, MF, \overrightarrow{fv} \rangle$. Specifically,

1) cv refers to the crisp value of the modeling entity at a certain time point.

2) *dom* is the domain of the crisp value of the entity.

3) $\overrightarrow{lv} = (lv_1, lv_2 ...)$ refers to the value of linguistic variable, e.g. (Middle, Low, Middle, High, Middle High). By a linguistic variable we mean a variable whose values are words or sentences in a natural or artificial language [24].

4) $MF = \{mf_1, mf_2, ...\}$ refers to the set of membership function and the relation between $lv_i$ and $mf_i$ is a Bijection.

5) $\overrightarrow{fv}$ is the vector of fuzzy value of $cv$ derived by MF where $fv_i = mf_i(cv)$.

We define the space of UE={$ue_1, ue_2...$} as the Cartesian product of all the available values of uncertain entities:

$$Space_{UE} \stackrel{\text{def}}{=} \otimes_{aue \in UE} dom(ue)$$

Based on the specified uncertain entities, we can easily describe uncertainties in context, tasks and softgoals of the mobile business application. The following specifications are the application on our example.

**Specification 1 (Uncertain Atomic Context)** Uncertain atomic contexts refer to the leaf nodes of the contextual model (Figure 2). These contexts are independent with each other. It is specified as $ac := \langle mv, dom, \overrightarrow{fv}, \overrightarrow{lv}, MF^{ac} \rangle$, where $mv$ is the monitored value of context which can be gauged by sensors. $dom$, $P_T$ and $P_E$ can be determined with the statistics of monitored value. MF can be determined with domain knowledge. For example, at a certain time point bandwidth rate (ac1) is specified as:

$$ac_1 := \langle 380kbps, [300, 500], (0.2, 0.8, 0), (Low, Middel, High), \{MF_1^{ac1}, MF_1^{ac1}, mf_3\} \rangle$$

$$MF_1^{ac1}(mv) = \begin{cases} \frac{400-mv}{100} & 300 \leq mv < 400 \\ 0 & 400 \leq mv \leq 500 \end{cases}, MF_2^{ac1}(mv) = \begin{cases} \frac{mv-300}{100} & 300 \leq mv < 400 \\ \frac{500-mv}{100} & 400 \leq mv \leq 500 \end{cases}, MF_3^{ac1}(mv) = \begin{cases} 0 & 300 \leq mv \leq 400 \\ \frac{mv-400}{100} & 400 < mv \leq 500 \end{cases}.$$

We define the context space as the Cartesian product of all the available monitored values of contexts:

$$Space_{context} \stackrel{\text{def}}{=} \otimes_{ac \in AC} dom(ac)$$

Where $AC=\{ac_1, ac_2...\}$.

**Specification 2 (Uncertain Task)** Uncertain tasks in our example is the leaf nodes of the goal model (Figure 1). An atomic task is specified as $t := \langle cp, dom, \overrightarrow{fv}, \overrightarrow{lv}, MF^t \rangle$, where $cp$ is the value of configuring parameter. *dom* is the system boundary of this task, which is determined by analyzers.

There are some differences from specifying atomic context. We distinguish the tasks derived by AND/OR-decompositions. To achieve g2, both t3 and t4 should be tuned during self-adaptation. Thus they both can serve as the conclusion in the reasoning rules. However, g1 should be accomplished either by configuring t1 or t2. If t1 and t2 both appear in the conclusion clause, the inconsistency between tasks will occur and the adaptation will not be achieved. Our solution is combining t1 and t2 as one variable in reasoning rules. Specifically, we assign one membership function for each task and the membership functions have no intersection. The linguistic value is name of task options. This solution can also be used for 3 or more optional tasks. Hence, the example specification of t1 and t2 goes like:

$$t_{1\&2} := \langle 15s, [-30, 30], (0, 0.5), (Network, GPS), \{MF_1^{t1\&2}, MF_2^{t1\&2}\} \rangle$$

$$MF_1^{t1\&2}(cp) = \begin{cases} \frac{cp+30}{15} & -30 \le cp \le -15 \\ \frac{-cp}{15} & -15 \le cp \le 0 \end{cases}, MF_2^{t1\&2}(cp) = \begin{cases} \frac{cp}{15} & 0 \le cp \le 15 \\ \frac{30-cp}{15} & 15 < cp \le 30 \end{cases}.$$

If the value is above zero, it means software uses GPS to locate users, otherwise it uses network. The specifications of *t3* and *t4* are similar to atomic context, so we do not repeat the specifying results here.

The configuration space is defined as the Cartesian product of all the available parametric values of tasks:

$$Space_{config} \stackrel{\text{def}}{=} \otimes_{t \in T} dom(t)$$

**Specification 3 (Uncertain Softgoal)** Softgoals are the modeling entities of NFRs. Except for the concerns of uncertainty, we also consider user preferences that are denoted by weights. A softgoal is specified as $sg := \langle sd, w, dom, \overrightarrow{fv}, \overrightarrow{lv}, MF^{sg} \rangle$, where *sd* is the value of satisfaction degree. In our study, we map *sd* to decimals in [0, 1] interval. 0 refers to extreme dissatisfaction, while 1 refers to extreme satisfaction. *w* refers to user preference of the softgoal, which is used for computing the trade-off decision during adaptation. The specification of *sg* is similar to atomic context and the satisfaction degree space is:

$$Space_{satDegree} \stackrel{\text{def}}{=} \otimes_{sg \in SG} dom(sg)$$

Where $SG=\{sg_1, sg_2...\}$. Apparently, each $dom(sg)$ is equal to [0, 1].

### 4.2 Specifying Relation

Uncertain relations are the basis of generating reasoning rules. It implies the dependency between entities. We define uncertain relations from the cause-and-effect viewpoint as follows:

**Definition 2 (Uncertain Relation)** An uncertain relation is a directed relation between two types of uncertain entities. It is specified as $UR := \langle \langle UE_1, UE_2 \rangle, W_{UR} \rangle$, in which $\langle UE_1, UE_2 \rangle$ is a pair of *UE* types, e.g. $\langle context, task \rangle$ and the direction is from $UE_1$ to $UE_2$. $W_{UR}$ is the weight matrix captures to what extent the elements in $UE_1$ effect the elements in $UE_2$. This relation maps the domain of $UE_1$ to the domain of $UE_2$ according to weight matrix. The mapping is defined as

$$\widetilde{\mathcal{M}_{UR}}: dom(UE_1) \xrightarrow{W_{UR}} dom(UE_2).$$

According to the discussed dependencies between uncertainties in Section 3, we consider three types of uncertain relations below. Each relation represents one of the three dependencies.

**Specification 4 (Evolve-Relation)** Evolve-relations capture the uncertain dependency from softgoals to contexts, which means the evolution of users' satisfaction degree depend on changes of contexts. It is specified as $EVO := \langle \langle AC, SG \rangle, W_{EVO} \rangle$. Figure 6 provides the graphical description of EVO-relation of our example. The weight matrix is elicited from the user preference questionnaire [31]. The positive weights mean that the increasing value of context variables will help the achievement of softgoals, while the negative weights mean the increasing value of context variables will hurt the achievement of softgoals. According to the definition of *UR*, the mapping is denoted as:

$$\widetilde{\mathcal{M}_{EVO}}: Space_{context} \xrightarrow{W_{EVO}} Space_{satDegree}.$$

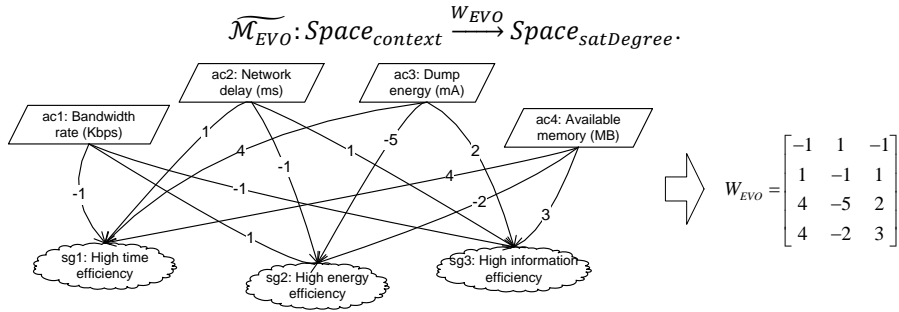

**Figure 5 Deriving weight matrix between atomic contexts and softgoals**

**Specification 5 (Satisfy-Relation)** Satisfy-relations describe the uncertain dependency from softgoals to tasks, which means the actual users' satisfaction degrees of softgoals depend on the changed task configurations. Similarly, it is specified as $SAT := \langle \langle T, SG \rangle, W_{SAT} \rangle$. The weight matrix is also elicited from the praference questionnaire. The mapping of SAT is defined as:

$$\widetilde{\mathcal{M}_{SAT}}: Space_{config} \xrightarrow{W_{SAT}} Space_{satDegree}$$

**Specification 6 (Adapt-Relation)** Adapt-relations describe the uncertain dependency from context to tasks, which means the changes of task configurations depend on contextual changes. It is specified as $ADP := \langle \langle AC, T \rangle, W_{ADP} \rangle$. The weight matrix is determined by requirements analyzers. The mapping of SAT is defined as:

$$\widetilde{\mathcal{M}_{ADP}}: Space_{context} \xrightarrow{W_{ADP}} Space_{config}$$

## 5. Reasoning rules

As we discussed early in the paper, due to the complexity of SAS, we can hardly build the analytic formulae the mapping problem proposed in Section 4. However, we can solve this problem by using heuristic reasoning. This reasoning process is based on user preference and analyzers' knowledge. In this section, we provide how to build these rules.

### 5.1 Specifying rules

To endow SAS with learning ability, we consider the software as either the gray-box system or the black-box system. Specifically, in gray-box systems, although the peculiarities of system internals are not entirely known, a certain model based on both insight into the system and experimental data can be constructed. This model, however, comes with a number of free parameters which can be estimated. In black-box systems, no prior model is available here, so everything has to be constructed from scratch, through observation and experimentation.

By considering SAS as a gray-box system, we build reasoning rules according to Mamdani fuzzy system [25], which is similar to the gray-box system. Because the outputs can be determined by identifying two kinds of parameters: linguistic variables and parameters of membership functions. In our work, linguistic variables are derived according to user preference, while parameters of membership functions need to be learned at runtime.

By considering SAS as a black-box system, we define reasoning rules according to T-S fuzzy system [26], which is similar to the black-box system, because the outputs are determined by identifying the analytic formulae of inputs. For the proposed problems, we can hardly identify the analytic formulae between the input variables and output variables. Thus, the analytic formulae can only be learned at runtime.

A Rule for gray-box reasoning is defined as

$$r^I: \text{IF } ue_1^1 \text{ is } lv_{r1}^{ue_1^1} \oplus \ldots \oplus ue_m^1 \text{ is } lv_{rm}^{ue_m^1} \text{ THEN } ue_1^2 \text{ is } lv_{s1}^{ue_1^2} \oplus \ldots \oplus ue_n^2 \text{ is } lv_{sn}^{ue_n^2}$$

Where $ue_i^j \in AUE_j$, $lv_r^{ue_i^j} \in lv^{ue_i^j}$ and $\oplus = \{\wedge, \vee\}$. For IF-clause, we can transform the expression into DNF and split it into several expressions of CNF. According to the specification of atomic tasks, the configurations of different tasks are decoupled, which makes the THEN-clause also a CNF. Hence, we simplify the rule as:

$$\wedge_{i=1}^m ue_i^1 \text{ is } lv_{ri}^{ue_i^1} \rightarrow \wedge_{i=1}^n ue_j^2 \text{ is } lv_{rj}^{ue_j^2}.$$

We denoted the rule set with $R_{UR}^I = \cup_{i=1}^n r_i^I$. For a given crisp vector of $UE_1$, the reasoning process to derive crisp vector of $UE_2$ with $\mathcal{R}^I$ is defined as: $\overrightarrow{cp^{UE_2}} = \widetilde{\mathcal{R}_{UR}^I}(R_{UR}^I, \overrightarrow{cp^{UE_1}}, MF^{UE_1} \cup MF^{UE_2})$. where $MF^{UE_1}$ is the set of membership functions of $UE_1$. $R_{UR}^I$ is used for building $\widetilde{\mathcal{M}_{EVO}}$, $\widetilde{\mathcal{M}_{SAT}}$ and $\widetilde{\mathcal{M}_{ADP}}$. To derive the membership function of $\widetilde{\mathcal{R}_{EVO}^I}$ and $\widetilde{\mathcal{R}_{SAT}^I}$, we capture the boundaries of linguistic variables from user preferences. For example, we design the question: 'What the value of dump energy is do you think is likely to be low, middle and high energy?' [27]. Then, we generate the triangle membership functions based on users' answers. Figure 7 presents a graphical illustration. For $\widetilde{\mathcal{M}_{ADP}}$, the membership functions is just initialed according to human knowledge at the beginning. However, the parameters of these functions will be revised during the learning process.

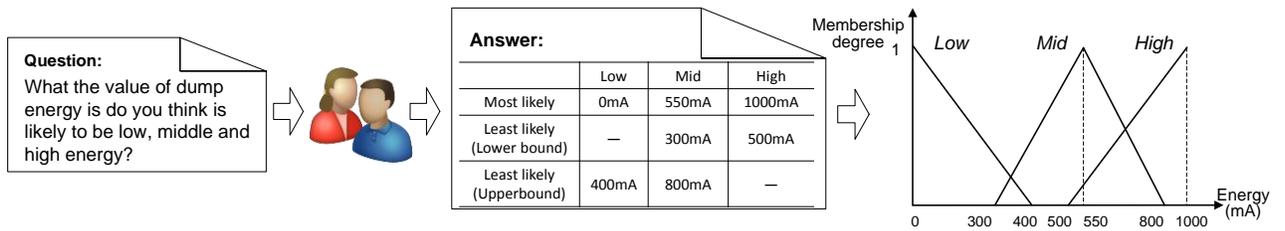

**Figure 6 Capturing membership function from stakeholders.**

A Rule for black-box reasoning is defined as

$$r^{II}: \text{IF } ue_1^1 \text{ is } lv_{r1}^{ue_1^1} \oplus \ldots \oplus ue_m^1 \text{ is } lv_{rm}^{ue_m^1} \text{ THEN } cp_j^2 = a_0 + \sum_{i=1}^m a_i * cp_i^1.$$

Where $cp_i^j$ is the crisp value of $aue_i^j$ and $a_i \in \mathbb{R}$. We simplify the rule as: $\wedge_{i=1}^m ue_i^1 \text{ is } lv_{ri}^{ue_i^1} \rightarrow cp_j^2 = a_0 + \sum_{i=1}^m a_i * cp_i^1$.

The rule set is formalized as $R_{UR}^{II} = \cup_{i=1}^n r_i^{II}$. Reasoning process with $\mathcal{R}^{II}$ is defined as $\overrightarrow{cp^{UE_2}} = \widetilde{\mathcal{R}_{UR}^{II}}(R_{UR}^{II}, \overrightarrow{cp^{UE_1}}, A)$ where $A$ is the matrix of coefficients. Different from $\widetilde{\mathcal{R}_{UR}^I}$, we do not need membership functions and linear parameters at the beginning. Each parameter of $\mathcal{R}^{II}$ can be learned. $\widetilde{\mathcal{R}_{UR}^{II}}$ is only used for reasoning on ADP-relation.

## 5.2 Generating Rules

According to the description above, to employ $R_{UR}^I$, we need to determine the linguistic vector $\overrightarrow{lv_{s1}^{ue_i^2}}$ in the THEN-clause. This algorithm is based on the elicited weight matrix $W_{UR}$ (Defination 3). For better understanding, we take EVO-relation as an example (Figure 6). Assume that each context and satisfaction degree is described with {*Low, Mid, High*}. All the possible combinations of linguistic values of contexts form a linguistic matrix *LM*. The steps are as follows:

1) Map *LM* to a numeric matrix *NM* according to the mapping: $f(Low)=1$, $f(Mid)=2$, $f(High)=3$.

2) Compute $NM*W_{EVO}$ and derive the boundary matrix *BM*. In our example (Figure 5), BM = $\begin{bmatrix} 6 & -23 & 3 \\ 26 & -5 & 17 \end{bmatrix}$.

3) For a given context vector $v=$(*Mid, Mid, High, Low*) = (2, 2, 3, 1), compute the score of the vector by $v* W_{EVO}$=(16,-17,9).

4) If we averagely trisect the first column of BM, i.e. [6, 26], the score 16 located in the interval [12.67, 19.33]. Then, the linguistic value of $sg_1$ should be Mid. Similarly, the linguistic values of $sg_2$ and $sg_3$ are Low and Low. Hence, the computed linguistic vector of THEN-clause is (*Mid, Low, Low*).

By iterating process 3 and 4, we can derive all the needed linguistic vectors of $R_{UR}^I$. After this step, we obtain all the conditions for reasoning with $R_{UR}^I$. The reasoning algorithms adopted is the classical algorithms on fuzzy sets [28]. Thus, we do not exhaust in this paper.

## 6. Adaptation mechanism

In this section, we first transform the problems brought out in Section 3 into the corresponding mathematical problems. Based on the specified targets and reasoning rules, we design four types of adaptation mechanism, each of which is underpinned by a reasoning schema.

### 6.1 Targets to Solve

Based on the specifications of uncertain attributes and reasoning rules, we specify the decision-making problem and the learning problem with corresponding expressions.

For a given monitor value $\overrightarrow{mv^{ac}}$ of contexts, the evolved satisfaction degree of softgoal is computed by

$$\overrightarrow{sd_{Desire}} = \widetilde{\mathcal{R}_{EVO}^I}(R_{EVO}^I, \overrightarrow{mv^{ac}}, MF^{AC} \cup MF^{SG})$$

Given a vector of task configurations $\overrightarrow{cp^t}$, the satisfaction degree of softgoal is derived with

$$\overrightarrow{sd_{Actual}} = \widetilde{\mathcal{R}_{SAT}^I}(R_{SAT}^I, \overrightarrow{cp^t}, MF^T \cup MF^{SG})$$

When contexts change, to achieve adaptation, a naive solution is reasoning with based on prior knowledge:

$$\overrightarrow{cp} = \widetilde{\mathcal{R}_{ADP}^I}(R_{ADP}^I, \overrightarrow{mv^{ac}}, (MF^{AC} \cup MF^T)_{prior})$$

The optimization solution to adaptation is

$$\overrightarrow{cp_{opt}} = argmin_{(\overrightarrow{cp} \in Space_{config})} \|\overrightarrow{sd_{Desire}} - \overrightarrow{sd_{Actual}}\|$$

Where $\|\overrightarrow{sd_{Desire}} - \overrightarrow{sd_{Actual}}\| = (\sum_{i=1}^{n}(sd_{Desire}^{sg_i} - sd_{Actual}^{sg_i})^2 \times w^{sg_i})/\sum_{i=1}^{n} w^{sg_i}$

The optimization solution to learning in gray-box system is

$$(MF^{AC} \cup MF^T)_{opt} = argmin_{(MF^{AC} \cup MF^T)} \|\overrightarrow{sd_{Desire}} - \widetilde{\mathcal{R}_{SAT}^I}(R_{SAT}^I, \widetilde{\mathcal{R}_{ADP}^I}(R_{ADP}^I, \overrightarrow{mv^{ac}}, MF^{AC} \cup MF^T), MF^T \cup MF^{SG})\|$$

Similarly, the optimization solution to learning in black-box system is

$$A_{opt} = argmin_{(A \in \mathbb{R})} \|\overrightarrow{sd_{Desire}} - \widetilde{\mathcal{R}_{SAT}^I}(R_{SAT}^I, \widetilde{\mathcal{R}_{ADP}^{II}}(R_{ADP}^{II}, \overrightarrow{mv^{ac}}, A), MF^T \cup MF^{SG})\|$$

After learning, SAS can used the new knowledge to configure the system by

$$\overrightarrow{cp} = \widetilde{\mathcal{R}_{ADP}^I}(R_{ADP}^I, \overrightarrow{mv^{ac}}, (MF^{AC} \cup MF^T)_{opt})$$

$$\text{And } \overrightarrow{cp} = \widetilde{\mathcal{R}_{ADP}^{II}}(R_{ADP}^{II}, \overrightarrow{mv^{ac}}, A_{opt})$$

To sum up, we derive four solutions to the adaptation and evolution problem: 1) naive adaptation, 2) optimized adaptation, 3) adaptation with learned knowledge in gray-box system and 4) adaptation with learned knowledge in black-box system.

### 6.2 Reasoning Schemas

To implement the above four solutions, we design four types of reasoning schemas from the control-theory perspective, including: forward reasoning (FR) schema, backward reasoning (BR) schema, parameter-identified reasoning (PR) schema and system-identified reasoning (SR) schema. Figure 8 depicts these reasoning schemas with inputs and outputs designated on the directed edges. The algorithms of schemas are presented in Figure 9.

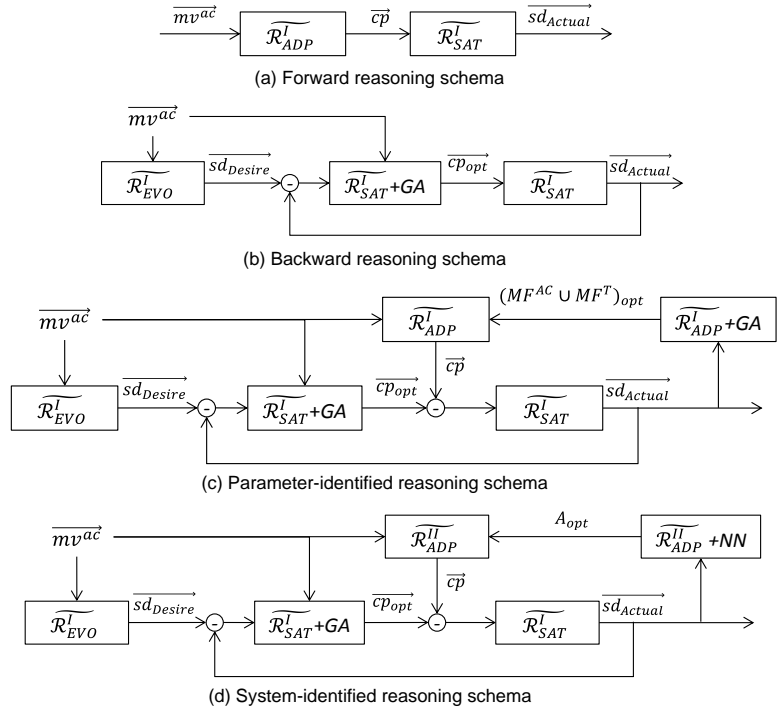

**Figure 7 Four types of reasoning schemas.**

**Forward Reasoning Schema**

FR is built based on the idea of feedforward control. Feedforward control measures the disturbances and adjusts the control inputs to reduce the impact of the disturbance on the system output. In our work, disturbances refer to contextual changes. Control inputs refer to task configurations, while the impact on SAS is considered as changes of NFRs. FR schema performs naive adaptation, which means the adaptation is achieved with fixed knowledge and does not consider whether the actual satisfaction degrees match the desired ones. When contexts change, this schema aims to directly tune the configurations.

**Backward Reasoning Schema**

BR schema is built based on the idea of feedback control. Feedback control loop is proven to be an appropriate way of building adaptation mechanisms in self-adaptive systems [29-32]. It aims to adjust the input according to the measured error and maintains the output sufficiently closed to what is desired. For our study, the outputs refer to the satisfaction degrees of NFRs, while the measured errors are the deviations between the desired satisfaction degrees and actual ones.

The algorithm depicts that before performing adaptation, SAS should know the changes in controlled variables (Algorithm 2 line 1), i.e. desired satisfaction degree of softgoals. The optimization is achieved through diminishing the deviations between desired and actual satisfaction degrees. This process is underpinned by GA.

**Parameter-identified Reasoning Schema**

PR schema treats SAS as the gray-box system and is designed based on the mechanism of feedforward-feedback control. Feedforward-feedback control mechanism has the advantage of both feedforward and feedback control schemes. First, it can tune system behavior based on the measured disturbances at runtime. Second, when deviations exist between the measured outputs and desired outputs, it can correct the behavior accordingly. It applies FR and BR as two loops to achieve adaptation. Firstly, it checks whether the knowledge in similar contexts can help self-adaptation through FR. If no appropriate knowledge is found, it turns to BR for achieving self-adaptation. The learning ability is implemented with $\widetilde{\mathcal{R}_{SAT}^{I}}$ and GA (Figure 8(c)). In GA, parameters of membership functions are coded into chromosomes. PR learns the optimal membership functions at each time point. Then, these functions are added into the knowledge base as revised knowledge. To test the performance of new knowledge, we use a set of thresholds to simulate the fault-tolerant scenarios (Algorithm 3 line 4). Obviously, the larger the threshold is, the better the performance will be.

**System-identified Reasoning Schema**

Similar with PR, SR schema is also based on the idea of feedforward-feedback control. However, it builds SAS as the black-box system. From Figure 8, we can find that the differences lie in the forward reasoning block and learning block. In forward reasoning block, PR uses the reasoning process $\widetilde{\mathcal{R}_{ADP}^{I}}$, while SR applies $\widetilde{\mathcal{R}_{ADP}^{II}}$. Besides, the learning process is based on $\widetilde{\mathcal{R}_{ADP}^{II}}$ and NNs. In Algorithm 4, BR is used for the early $T$ time steps, because SAS has no available prior knowledge and the learning process of NNs needs to collect a number of data pairs, i.e. $(\overrightarrow{mv^{ac}}, \overrightarrow{mv^t})$. For increasing the precision of learning, we classify the emerged contexts with the Fuzzy c-Means method (Algorithm 4 line 13). For each context class, NNs are applied to learning the optimal coefficient matrix $A_{opt}$ of $\mathcal{R}^{II}$.

---

**Algorithm 1** Forward Reasoning

**Input:** $R_{ADP}^I, \overrightarrow{mv^{ac}}, MF^{AC} \cup MF^T, R_{SAT}^I, \overrightarrow{cp^{AT}}, MF^T \cup MF^{SG}$
**Output:** $\overrightarrow{cp}, \overrightarrow{sd_{Actual}}$

1:  $\overrightarrow{cp_{actual}} := \widetilde{\mathcal{R}_{ADP}^I}(R_{ADP}^I, \overrightarrow{mv^{ac}}, MF^{AC} \cup MF^T)$
2:  $\overrightarrow{sd_{Actual}} := \widetilde{\mathcal{R}_{SAT}^I}(R_{SAT}^I, \overrightarrow{cp_{actual}}, MF^T \cup MF^{SG})$
3:  $\overrightarrow{cp} := \overrightarrow{cp_{actual}}$
4:  return $\overrightarrow{cp}, \overrightarrow{sd_{Actual}}$

---

**Algorithm 2** Backward Reasoning

**Input:** $R_{EVO}^I, \overrightarrow{mv^{ac}}, MF^{AC} \cup MF^{SG}, R_{SAT}^I, MF^T \cup MF^{SG}$, N, MAXgen
**Output:** $\overrightarrow{cp_{opt}}, \overrightarrow{sd_{Actual}}$

1:  $\overrightarrow{sd_{Desire}} := \widetilde{\mathcal{R}_{EVO}^I}(R_{EVO}^I, \overrightarrow{mv^{ac}}, MF^{AC} \cup MF^{SG})$
2:  chrom := createPopulation(N, $Space_{config}$)
3:  **for all** $\overrightarrow{chr} \in chrom$
4:      $\overrightarrow{sd_{Actual}} := \widetilde{\mathcal{R}_{SAT}^I}(R_{SAT}^I, \overrightarrow{cp}, MF^T \cup MF^{SG})$
5:  **end for**
6:  **while** generation < MAXgen
7:      perform GA
8:      $\overrightarrow{cp_{opt}} := argmin_{(\overrightarrow{chr} \in Space_{config})} \|\overrightarrow{sd_{Desire}} - \overrightarrow{sd_{Actual}}\|$
9:  **end while**
10: $\overrightarrow{sd_{Actual}} := \widetilde{\mathcal{R}_{SAT}^I}(R_{SAT}^I, \overrightarrow{cp_{opt}}, MF^T \cup MF^{SG})$
11: return $\overrightarrow{cp_{opt}}, \overrightarrow{sd_{Actual}}$

---

**Algorithm 3** Parameter-identified Reasoning

**Input:** $R_{EVO}^I, \overrightarrow{mv^{ac}}, MF^{AC} \cup MF^{SG}, R_{SAT}^I, MF^T \cup MF^{SG}$,
   $R_{ADP}^I, MF^{AC} \cup MF^T$, N, MAXgen, Threshold
**Output:** $\overrightarrow{cp_{opt}}, \overrightarrow{sd_{Actual}}, (MF^{AC} \cup MF^T)_{opt}$

1:  $\overrightarrow{sd_{Desire}} := \widetilde{\mathcal{R}_{EVO}^I}(R_{EVO}^I, \overrightarrow{mv^{ac}}, MF^{AC} \cup MF^{SG})$
2:  $\overrightarrow{mv_{nearest}^{ac}} := findNearst(AC)$
3:  $[\overrightarrow{cp_{actual}}, \overrightarrow{sd_{Actual}}] := ForwardReasoning(\ )$
4:  **if** $\|\overrightarrow{sd_{Desire}} - \overrightarrow{sd_{Actual}}\| \leq$ Threshold
5:      $\overrightarrow{cp_{opt}} := \overrightarrow{cp_{actual}}$
6:      $(MF^{AC} \cup MF^T)_{opt} := (MF^{AC} \cup MF^T)_{nearest}$
7:  **else** $[\overrightarrow{cp_{opt}}, \overrightarrow{sd_{Actual}}] := BackwardReasoning(\ )$
8:      MFchrom := createPopulation(N, $Space_{config}$)
9:      **for all** $\overrightarrow{MFchr} \in MFchrom$
10:         $\overrightarrow{cp_{actual}} := \widetilde{\mathcal{R}_{ADP}^I}(R_{ADP}^I, \overrightarrow{mv^{ac}}, \overrightarrow{MFchr})$
11:     **end for**
12:     **while** generation < MAXgen
13:         perform GA
14:         $(MF^{AC} \cup MF^T)_{opt} :=$
               $argmin_{(\overrightarrow{MFchr})} \|\overrightarrow{sd_{Desire}} - \overrightarrow{sd_{Actual}}\|$
15:     **end while**
16: **end if**
17: return $\overrightarrow{cp_{opt}}, \overrightarrow{sd_{Actual}}, (MF^{AC} \cup MF^T)_{opt}$

---

**Algorithm 4** System-identified Reasoning

**Input:** $R_{EVO}^I, \overrightarrow{mv^{ac}}, MF^{AC} \cup MF^{SG}, R_{SAT}^I, MF^T \cup MF^{SG}$,
   $R_{ADP}^{II}$, Threshold, T
**Output:** $\overrightarrow{cp_{opt}}, \overrightarrow{sd_{Actual}}, A_{opt}$

1:  $\overrightarrow{sd_{Desire}} := \widetilde{\mathcal{R}_{EVO}^I}(R_{EVO}^I, \overrightarrow{mv^{ac}}, MF^{AC} \cup MF^{SG})$
2:  **if** time$\leq T$
3:      $[\overrightarrow{cp_{opt}}, \overrightarrow{sd_{Actual}}] := BackwardReasoning()$
4:      $A_{opt} := NN(mv^{AC}, cp^T)$
5:  **else**
6:      $\overrightarrow{mv_{nearest}^{ac}} := findNearst(AC)$
7:      $[\overrightarrow{cp_{actual}}, \overrightarrow{sd_{Actual}}] := ForwardReasoning(R_{ADP}^{II}, A_{nearest})$
8:      **if** $\|\overrightarrow{sd_{Desire}} - \overrightarrow{sd_{Actual}}\| \leq$ Threshold
9:          $\overrightarrow{cp_{opt}} := \overrightarrow{cp_{actual}}$
10:         $A_{opt} := A_{nearest}$
11:     **else** $[\overrightarrow{cp_{opt}}, \overrightarrow{sd_{Actual}}] := BackwardReasoning(\ )$
12:         K:=time%T
13:         contextClass:=FCM($mv^{AC}$, K)
14:         **for all** cla $\in$ contextClass
15:             $A_{cla,opt} := NN(mv_{cla}^{AC}, cp_{cla}^{AC})$
16:         **end for**
15:     **end if**
16: **end if**
17: return $\overrightarrow{cp_{opt}}, \overrightarrow{sd_{Actual}}, A_{opt}$

**Figure 8** Algorithms of reasoning schemas

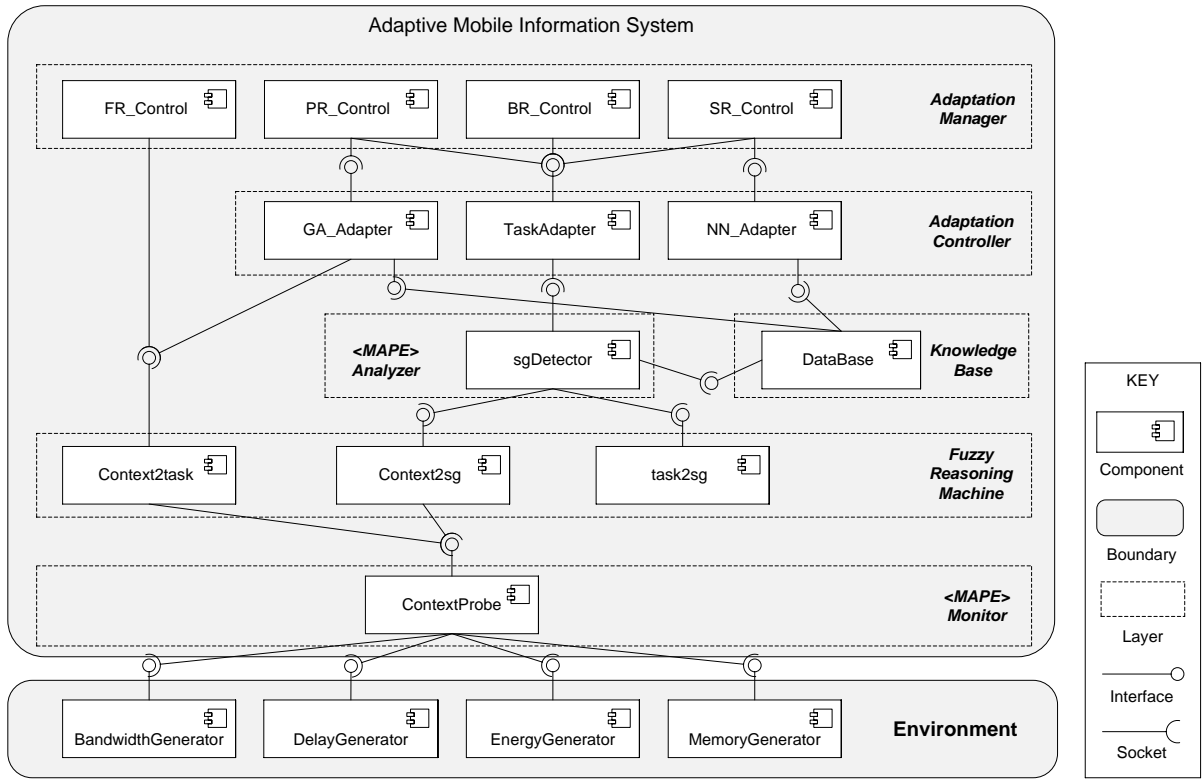

## 7. Evaluation

This section presents the evaluation of the proposed approach on the mobile business application. Firstly, we introduce the design of our simulative experiment, including the experiment processes and parameters used. Then, we elaborate the synthesized results and implied conclusions.

### 7.1 Design and Settings

To evaluate the performance of our approach with contextual changes, we implement each schema in three types of contexts: derivable context, quasi-noisy context and noisy context. The derivable context refers to contextual changes are subjected to a derivable function. It is used to test the approach with continuously changing contexts. Quasi-noisy context is generated by adding slight gauss noises to the derivable context. It is used to imitate the contexts which change with some trends. Noisy context refers to the randomly changed contexts. Figure 9 presents the contextual data used by our approach. Besides, for illustrating the advantages of learning and evolution, we use a set of thresholds to imitate the fault-tolerant situations.

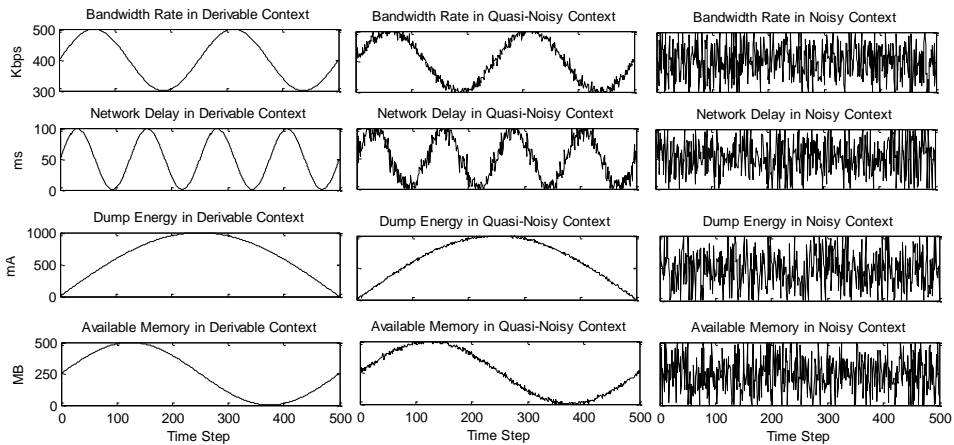

**Figure 9 Three types of contexts used.**

Except for contextual data, the settings of membership functions and linguistic vectors of reasoning rules are derived based on the steps provided in Section 5. Each context, task and softgoal is described with three linguistic terms. The settings of GA and NNs are provided in Table 1. All the experiments reported were performed on a machine with an AMD FX8350 (3.8GHz) CUP with 4GB of RAM.

**Table 1 Setting of GA and NNs**

| Algorithm | Settings |
|---|---|
| GA | chromLength=20, populationSize=100, maxGen=100, genGap=0.9, crossoverRate=0.9, mutRate=0.05, selMethod='stochastic universal sampling', crosoverMethod='Single-point crossover', |
| NNs | epochs=100, goal=0.05, rate=0.05 |

### 7.2 Results and conclusion
**Adaptation results**

Figure 10 presents the evolved satisfaction degrees of softgoals of the mobile business application. These results illustrate that our approach of modeling and reasoning over uncertainties can reflect users' preferences correctly. For example, we can find that in derivable contexts, satisfaction degree of energy efficiency is highly correlated to dump energy, because the EVO-relation between energy efficiency and dump energy has higher weight (Figure 5).

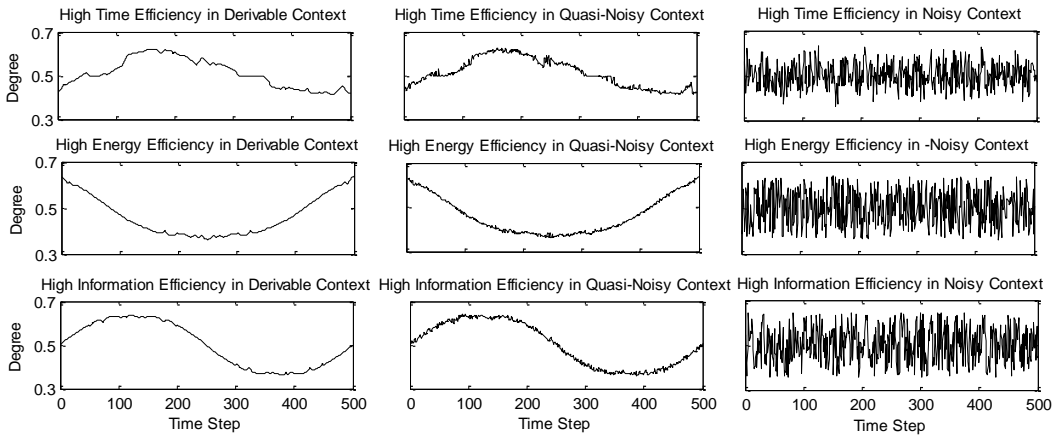

**Figure 10 Evolved satisfaction degrees of softgoals**

The optimal configurations derived through BR are provided in Figure 11. From the former two columns, it depicts that the approach is insensitive to less noises. Besides, the results also reflect users's preferences. For example, the satisfaction degree of information efficiency is highly correlated to the configuration of data size and the satisfaction degree of time efficiency is highly correlated to the configuration of time interval. These results present the availability of our approach in solving the decision-making problem and derive the optimal configurations.

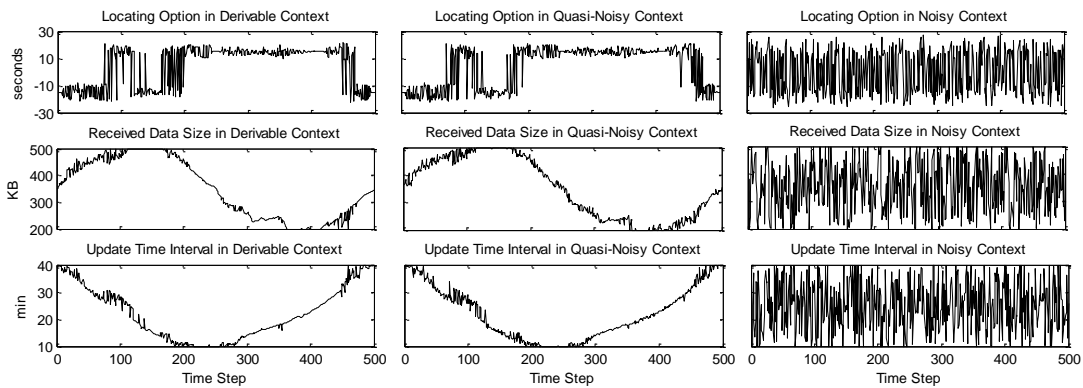

**Figure 11 Optimal configurations derived through BR.**

Figure 12 presents the number of time steps when the system succeeds in adaptation with learning knowledge by applying PR and SR. It shows that when the tolerant deviation gets larger, the learned knowledge provides better adaptation performance. Besides the learned knowledge make PR more flexible than SR. It is because PR considers SAS as gray-box system and the evolution demands learning partial parameters, i.e. parameters of membership functions. It also implies that identifying analytic formulae of SAS is not the best choice for building adaptation mechanisms.

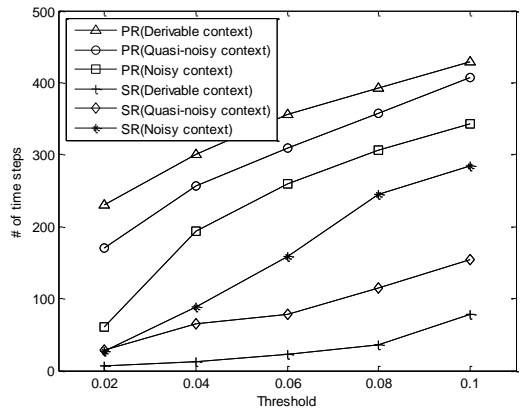

**Figure 12 Performance of learned knowledge in PR and SR**

### Performance analysis

We focus on two kinds of performances: accuracy (of satisfaction degrees) and execution time. Figure 13 presents the deviations between desired satisfaction degrees and actual ones by applying FR and BR, which is measured by weighted Euclidean Distance (Section 6.1). Each box contains the results of 500 time steps. It depicts that, to achieve self-adaptation, BR performs much better than FR. It also reveals that analyzers' assumption towards a system may deviate from stakeholders' expectations. This is the reason why we need to endow SAS the ability of learning and evolving at runtime. FR costs about 0.003s for each adaptation, while BR costs about 0.35s. Thus, FR is much faster than BR.

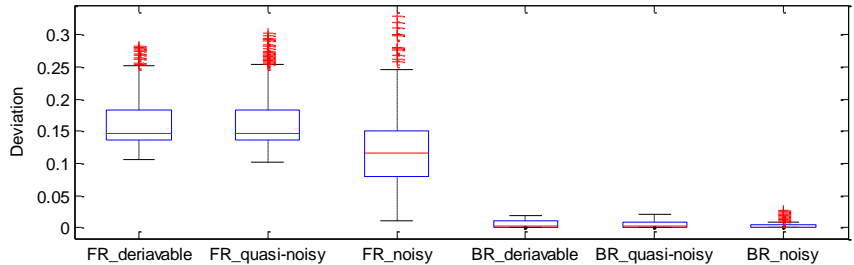

**Figure 13 Deviations by using FR and BR**

Figure 14 presents the deviations between desired satisfaction degrees and actual ones by using PR and SR. The discrete data depict that though adaptation can be achieved, both schemas are unstable in noisy contexts. For PR-based adaptation, as the threshold gets larger, the number of discrete points gets increased. These results provide the choice for build adaptation mechanisms under different contexts and with different fault-tolerant qualities. For example, it implies that if the system can tolerate 8% deviations in quasi-noisy context, we prefer to use SR to achieve adaptation and evolution, because the deviation is more stable. The time cost by PR is about 0.22s and the time cost of SR is about 0.006s.

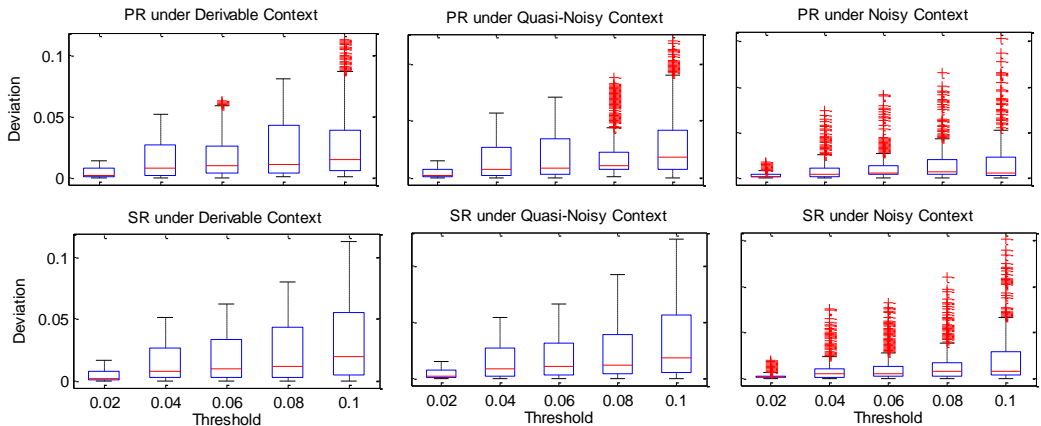

**Figure 14 Deviations by using PR and SR**

Based on these results, we conclude the comparison in Table 2, which can be used to guide requirements analyzers to build appropriate adaptation mechanisms in different contexts with our approach.

**Table 2 Comparison results of different reasoning schemas**

| Schema | Adaptation (Optimizing) | Evolution (Learning) | Cost | Appropriate context |
|---|---|---|---|---|
| FR | no | no | least | no appropriate context |
| BR | yes | no | most | noisy context |
| PR | yes | yes | more | derivable and quasi-noisy context (low fault-tolerance) |
| SR | yes | yes | less | derivable and quasi-noisy context (high fault-tolerance) |

## 8. Related work

**Dealing with uncertainty**. The concept of uncertainty was elaborated in pioneering works [6-8]. Sawyer *et al*. [8] provided a research agenda for dealing with requirements uncertainties. They argue requirements for self-adaptive systems should be viewed as runtime entities that can be reasoned over in order to understand the extent to which they are being satisfied and to support adaptation decisions that can take advantage of the systems' self-adaptive machinery. Ramirez *et al*. [7] introduced the definition and taxonomy of uncertainty in the context of dynamically adaptive systems and identified existing techniques for mitigating different types of uncertainty. More recently, in roadmap paper [6], Esfahani and Malek characterized sources of uncertainty in SAS and discussed the state-of-the-art for dealing with uncertainty. Baresi *et al*. [9] proposed FLAGS for mitigating requirements uncertainty by extending goal model with adaptive and fuzzy goals. With this approach, requirements can be partially satisfied and the system possesses the ability of fault tolerance. Whittle *et al*. [10] proposed RELAX, a formal requirements specification language, for specifying the uncertain requirements in SAS. With RELAX, we can establish the boundaries of adaptive behavior. In their following work [33], Cheng *et al*. introduced a goal-based modeling approach to development requirements for dynamically adaptive systems when identifying uncertainty factors in the environment. Different from their works, we not only describe the fuzziness of requirements but also take context uncertainties and configuration uncertainties into consideration. Moreover, our specification can be used to generate reasoning rules and implement quantitative reasoning. FUSION was proposed by Elkhodary *et al*. [11]. The approach uses online learning to mitigate the uncertainty associated with changes in context and tune system behaviors to unanticipated changes. Esfahani *et al*. [12] proposed POISED for improving the quality attributes and achieve a global optimal configuration of a system by assessing both the positive and negative consequences of context uncertainty. However, their approaches are proposed based on existing configuration options, i.e. they only deal with structural adaptation. In our approach, we consider both structural adaptation and parametric adaptation. Besides, our approach implements decision-making through reasoning, while their approaches achieve adaptation through the given linear functions.

**Building adaptation mechanism**. Brun *et al*. [29] explored and elaborated how feedback loops can be utilized in engineering self-adaptive systems, especially the MAPE loop [34], which includes monitoring, analyzing, planning and executing processes. The MAPE loop is a wildly used feedback loop for building adaptation mechanism in SAS. Wang *et al*. [35] focused on monitoring and analysis aspect. They proposed a framework for diagnosing failure of software requirements by transforming the diagnostic problem into a propositional satisfiability problem, which is solved by SAT solvers. In [36], Wang and Mylopoulos proposed an autonomic architecture consisting of monitoring, diagnosing, reconfiguration and execution component. This architecture uses requirements models as a basis for monitoring, diagnosis and reconfiguration. The monitoring component monitors the satisfaction of software requirements and generates log data. When errors are found, the diagnostic component infers the denial of the requirements and identifies problematic components. For repair, the reconfiguration component selects a best system reconfiguration among all competing reconfigurations that are free of failures. The chosen reconfiguration contributes most positively to the systems NFRs and it minimally reconfigures the system from its current configuration. The execution component runs compensation actions to restore the system to its previous consistent state, and reconfigures the system under the chosen configuration. Similar to their work, our adaptation schemas also implied the MAPE process. By monitoring, SAS gauges the values of contextual variables and computes the evolved requirements satisfaction degrees. When deviations of satisfaction degrees are found, SAS will perform the reasoning schemas to repair the deviations. Then, SAS chooses the available configuration options and tunes the task parameters. Differently, our approach supports tunes the configuration parameters continuously and supports reasoning with uncertainties.

## 9. Conclusion and future Work

In this paper, we proposed an approach to specifying uncertainties, achieving self-adaptation and evolution for self-adaptive software. It uses fuzzy logic to capture the ambiguities of contexts, system tasks and NFRs, which are considered as core concepts in goal-oriented requirements engineering. Based on the description of fuzziness, we generate a series of reasoning rules to build the computational relationship between these core concepts. To endowing the ability of adaptation and evolution, we proposed four types of adaptation mechanism, which can be implemented through fuzzy reasoning and supervised learning. Our approach has been applied to a mobile business application and the results imply meaningful conclusions.

The approach can serve as a general guidance for modeling uncertainties at RE stage, especially after eliciting users' preferences. It can be used to check whether stakeholders' expectations are reasonable for the system to be developed and whether system analyzers' knowledge is consistent with these expectations. The enriched knowledge can help analyzers have a better understanding about system and guide them to build more appropriate quantitative relations between contexts and SAS. Moreover, the approach can be used to solve the adaptation problem of continuously changed contexts. This provides more precise analysis of system behavior than analyzing discrete task options.

Future work includes two lines. First, we concern the other type of uncertainties, i.e. probability. Different from fuzziness, probability captures variability and describes the dynamic attributes of systems. However, we can only handle a finite amount of probabilities, which demands us to split system behavior into several states. For SAS whose task parameters can be tuned continuously, e.g. the mobile business application, splitting behavior can be performed by classifying the task parameters. This process can also be used to classify contexts. Hence, we can model contexts and SAS as a Hidden Markov Model, in which system states are hidden states and contextual states are observed state. Based on transition probabilities between system states and emission probability between contexts and system states, for given contexts, we can predict what will be the probable next system state. This process may efficiently reduce the solution space in our approach. We also concern the verification with uncertain attributes, especially the flexibility provided by introducing fuzziness into requirements. To this end, we need to build effective adaptation loops, e.g. MAPE loops and reasonably describe contextual changes, fuzzy requirements and adaptation logic. We consider UPPAAL-SMC as a proper solution to this problem.